\documentclass[twocolumn,preprintnumbers,amsmath,amsfonts,amssymb,floatfix,aps,pra,superscriptaddress]{revtex4}

\usepackage[utf8]{inputenc}

\usepackage{graphicx}
\usepackage{amssymb}
\usepackage{amsmath}
\usepackage{amsfonts}
\usepackage{bm}
\usepackage{color}
\usepackage{hyperref}

\newcommand{\op}[1]{\hat{#1}}

\newcommand{\be}{\begin{equation}}
\newcommand{\ee}{\end{equation}}
\newcommand{\idk}{\int\hspace{-.2cm}\frac{d^3k}{(2\pi)^3}}

\begin{document}
\title{Mean-field vs RPA calculation of the energy of an impurity immersed in a spin 1/2 superfluid}

\author{A. Bigu\'e}%
\email{Corresponding author: arnaud.bigue@phys.ens.fr}
\affiliation{Laboratoire de physique de l'Ecole Normale sup\'erieure, ENS, Universit\'e PSL, CNRS, Sorbonne Universit\'e, Universit\'e de Paris, F-75005 Paris, France.}

\author{F. Chevy}%
\affiliation{Laboratoire de physique de l'Ecole Normale sup\'erieure, ENS, Universit\'e PSL, CNRS, Sorbonne Universit\'e, Universit\'e de Paris, F-75005 Paris, France.
}%

\author{X. Leyronas}%
\affiliation{Laboratoire de physique de l'Ecole Normale sup\'erieure, ENS, Universit\'e PSL, CNRS, Sorbonne Universit\'e, Universit\'e de Paris, F-75005 Paris, France.
}%


\begin{abstract}
In this article we calculate the energy of an impurity weakly coupled to a spin 1/2 fermionic superfluid. We show that the divergences resulting from three-body physics can only be cured using a proper description of the excitations of the many-body background. We highlight the crucial role played by interactions between quasiparticles which are  overlooked within BCS (Bardeen-Cooper-Schrieffer) mean-field theory of fermionic superfluidity. By contrast, we prove that their addition using the  Random Phase Approximation (RPA) allows us to regularize  the energy of the impurity. Finally, we show that these beyond mean-field corrections should be observable by the analysis of the frequency shift of the impurity center of mass oscillations in an external confining potential.
\end{abstract}

\maketitle

\section{Introduction and main results}

The concept of quasi-particle provides a simple and powerful framework for the study of the low-energy physics of complex many-body systems. In practice, a quasi-particle can be described as a free particle dressed by a cloud of excitations of the surrounding many-body medium leading to a renormalization of   its physical properties, and most notably its mass.

This mechanism was first suggested by Landau and Pekar to describe the coupling of electrons to the vibration modes of a crystal lattice (the so-called polaron problem)  \cite{landau1948effective} and since then has been generalized to a host of physical situations, from solid state physics
 where polarons play a crucial role in the study a photovoltaic materials \cite{alexandrov2008polarons}, to nuclear physics \cite{zuo20041s0}.

In the recent years, mixtures of ultracold atoms have provided an ideal playground for the study of polaron physics \cite{massignan2014polarons}. In this context, two simple systems were studied. Firstly, an analogous of Landau-Pekar polaron was obtained by immersing an impurity in a weakly interacting Bose-Einstein condensate (BEC) - the so-called Bose-polaron  \cite{jorgensen2016observation,hu2016bose,levinsen2015impurity}. In this case, the bogoliubov spectrum describing the low-lying excitations of the BEC possess a structure that is similar to the phonons of a crystal. Another simple situation was obtained by considering the case of a particle swimming in a sea of non-interacting spin-polarized fermions, the so-called Fermi polaron \cite{chevy2006upa,lobo2006nsp,prokof'ev08fpb,nascimbene2009pol,schirotzek2009ofp,yan2019boiling,mathy2011} that was also realized in exciton-polariton systems \cite{sidler2017fermi}.

A generalization of these two polaronic model systems is provided by the study of an impurity immersed in an ensemble of attractive spin 1/2 fermions \cite{Ferrier2014Mixture,roy2017two,yao2016observation}.
When the attraction between the particles of the medium is varied, the ground state of the many-body background evolves from an ideal gas of fermions to a Bose-Einstein condensate of strongly bound dimers, thus realizing the celebrated BEC-BCS crossover \cite{zwerger2012BCSBEC}. As a consequence, the state describing the impurity immersed in a fermionic superfluid interpolates between the Fermi and Bose-polarons. While in reported experiments, the impurity is weakly coupled to the medium and most can be captured quantitatively by treating the impurity-fermion interaction within a mean-field approximation, a theoretical study of beyond-mean field effects  was initiated in \cite{yi2015polarons,nishida2015polaronic,Pierce19few} that highlighted the role of three-body interactions. In particular, \cite{Pierce19few} showed that the leading order corrections were related to the compressibility of the background medium after a regularisation of UV divergences made possible by the introduction of  explicit 3-body interactions \cite{hammer2015three}.

Building on this work, we show here that a precise description of the properties  of the background is required for the regularization procedure used in \cite{Pierce19few} to be effective. Indeed,  describing the background superfluid within BCS approximation that considers only pair-breaking excitations is incompatible with this renormalization scheme. To obtain a finite beyond-mean field correction, we need to take into account collective modes of the system, and for this we work within the framework of the  Random Phase Approximation (RPA) \cite{minguzzi2001dynamic,Combescot2006,kurkjiantempereklimin2020} that allows us to make quantitative predictions for the energy of the impurity.

More precisely, consider  an impurity of mass $m_{\rm i}$ immersed in a many-body ensemble of spin 1/2 fermions of mass $m_{\rm f}$.  $a$ and $a'$ are respectively the fermion-fermion and impurity-fermion scattering lengths (the latter being assumed   to be spin independent). We note $|\alpha\rangle$ the eigenstates of the medium in the absence of impurity, and $E_\alpha$ the corresponding eigenenergies. By convention, $\alpha=0$ corresponds to the ground state. For a total density $n$ of fermions, the state of the many-body background is characterized by the dimensionless parameter $1/k_F a$, where $k_F=(3\pi^2 n)^{1/3}$ is the Fermi wavector, the limits $1/k_F a\rightarrow -\infty$ (resp. $+\infty$) corresponding respectively to the weakly (resp. strongly) attractive regimes.

In the quasi-particle picture, the low-lying energy of an impurity of momentum $\hbar\bm q$ takes the form of a free particle dispersion relation
\be
E(\bm q)=E_0+\Delta E+\frac{\hbar^2 q^2}{2m^*}+o(q^2).
\ee
where $\Delta E$ is the interaction energy of the impurity with the many-body ensemble and $m^*$ is its effective mass.

Using perturbation theory, we have at second order in impurity-fermion coupling \cite{Pierce19few}

\begin{eqnarray}
 \frac{1}{m^*}&=&\frac{1}{m_{\rm i}}\left[1-\frac{4 g'^2n}{3}\int\frac{d^3\bm q}{(2\pi)^3}\varepsilon_q^{(\rm i)}\chi_3(\bm q,\varepsilon_q^{(\rm i)})\right]\label{Eq:EffectiveMass}\\
\Delta E&=&g'n\left[1+g'\int\frac{d^3\bm q}{(2\pi)^3}\left(\frac{1}{\varepsilon_q^{(\rm r)}}-\chi_1(\bm q,\varepsilon_q^{(\rm i)})\right)\right]\label{Eq:InteractionEnergy}
\end{eqnarray}
Here, $g'=2\pi \hbar^2 a'/m_{\rm r}$ is the  coupling constant of the impurity-medium two-body contact interaction, $m_{\rm r}$ is the reduced mass of an impurity-fermion pair, $\varepsilon_q^{(\alpha={\rm i},{\rm r})}=\hbar^2 q^2/2m_\alpha$ and

\be
\chi_p(\bm q,E)=\frac{1}{N}\sum_{\alpha}\frac{\left|\langle \alpha|\hat{n}_{{\bf q}}|0\rangle\right|^2}{(E+E_\alpha-E_0)^p}=\frac{(-1)^{p-1}}{(p-1)!}\frac{\partial^{p-1}\chi_1}{\partial E^{p-1}},
\label{eqdefchiRqE}
\ee
where $\hat{n}_{{\bf q}}=\sum_{{\bf k},\sigma}c^{\dagger}_{{\bf k},\sigma}c^{\phantom{}}_{{\bf k}+{\bf q},\sigma}$ is the Fourier transform of  the density operator and $N$ is the total number of fermions. $\chi_p$ plays a central role in the following and we note that it depends only on the properties of the excitation spectrum  of the many-body background and can be related to the density-density response function (see Sec. \ref{sec:chi}).

$\op n_{\bm q}$ is an operator that transfers a momentum $-\hbar\bm q$ to the many-body system. For large $\bm q$, we can assume that it couples the ground-state to free particle excitations for which $E_\alpha-E_0\simeq \hbar^2 q^2/2m_{\rm f}$. We therefore have
\be
\chi_p(q,\varepsilon_q^{(\rm i)})\underset{q\rightarrow\infty}{\sim} \frac{1}{N}\sum_{\alpha}\frac{\left|\langle \alpha|\op n_{\bm q}|0\rangle\right|^2}{\left(\varepsilon_q^{(\rm r)}\right)^p}\sim \frac{1}{\left(\varepsilon_q^{(\rm r)}\right)^p}.
\label{Eq:HighQExpansion}
\ee

As a consequence, we readily conclude that the sum appearing in the expression of the effective mass (Eq. (\ref{Eq:EffectiveMass})) is convergent and does not need any regularization. By contrast, the case of the interaction energy is more involved. Indeed, although $\chi_1$ and $1/\varepsilon^{(\rm r)}$ compensate at leading order in Eq.  (\ref{Eq:InteractionEnergy}), it was shown in \cite{Pierce19few} that the sum is  log-divergent as a consequence of the following large momentum behaviour of $\chi_1$

\be
\chi_1(q,\varepsilon^{\rm (i)}_q)\underset{q\rightarrow\infty}{=}\frac{1}{\varepsilon^{\rm (r)}_q}\left[1-\pi^2\kappa(\eta)\frac{m_{\rm f}}{m_{\rm r}}\frac{C_2}{Nq}+...\right],
\label{Eq:AsymptoticChi}
\ee
where $C_2$ is Tan's contact parameter of the many-body background \cite{tan2008large}, $\eta=m_{\rm i}/m_{\rm f}$ and

\be
\begin{split}
\kappa(\eta)=&\frac{\sqrt{\eta^3 (\eta+2)}}{2\pi^3  (\eta+1)^2}-\frac{\eta}{2\pi^3 }\arctan
   \left(\frac{1}{\sqrt{\eta (\eta+2)}}\right)\\
   &-\frac{4 }{\pi^3}  \sqrt{\frac{\eta}{ \eta+2}}\arctan
   \left(\sqrt{\frac{\eta}{\eta+2}}\right)^2.
\end{split}
\label{eq:Kappa}
\ee

Following \cite{hammer2015three}, this divergence can be regularized using an effective three-body interaction leading to the following regularized expression for the polaron energy shift
\be
\begin{split}
\Delta E=g'  n\biggl[ 1+&k_F a' F\left(\frac{1}{k_F a}\right)\\
&-2\pi\frac{m_{\rm f}}{m_{\rm r}}\kappa(\eta)\frac{a'C_2}{N}\ln(k_FR_3)+...\biggr],
\end{split}
\label{eq:Born2}
\ee
where $R_3$ is an effective length characterizing the three-body scattering amplitude  ($R_3/a'\simeq 1.5$ for the Lithium 6/Lithium 7 mixture used in \cite{Ferrier2014Mixture}) and

\be
\begin{split}
F\left(\frac{1}{k_Fa}\right)\underset{\Lambda \rightarrow \infty}{=}\frac{2\pi}{k_F}\biggl[\frac{\hbar^2}{m_{\rm r}}\int_{q<\Lambda}&\frac{d^3\bm q}{(2\pi)^3}\left(\frac{1}{\varepsilon^{\rm (r)}_q}-\chi_1(q,\varepsilon^{(\rm i)}_q)\right)\\
&-\frac{m_{\rm f}}{m_{\rm r}}\kappa(\eta)\frac{C_2}{N}\ln(\Lambda/k_F)\biggr]
    \end{split}
\label{Eq:MainEq}
\ee

To calculate $\chi_p$, a first approach is to use the extension of the BCS theory-a mean-field theory- to the whole BEC-BCS crossover \cite{Pierce19few}. In this case, we readily obtain

\be
\chi_{\rm MF,\,p}(q,E)=\frac{1}{n}\idk\frac{\left |u_{\bm k}v_{\bm k+\bm q}+v_{\bm k}u_{\bm k+\bm q}\right |^2}{\left(E+ E_{\bm k}+E_{\bm k+\bm q}\right)^p},
\label{Eq:chipMF}
\ee
where $(u_k,v_k)$ are the amplitudes of the Bogoliubov modes and $E_k$ their energy \cite{de2018superconductivity}.

However, this type of approximation scheme is restricted to the pair-breaking excitations sector and does not account for collective modes (phonons) that are dominant on the BEC side of the crossover. A well known consequence of this peculiarity is for instance that the compressibility calculated using BCS theory does not satisfy the f-sum rule. In our case, this leads to an underestimated value of $\kappa$ (the third term appearing in Eq. (\ref{eq:Kappa}) is missing, see sec. \ref{sec:highq}) which leads to a UV-divergence of $F$ when $\Lambda\rightarrow\infty$ and prevents the regularization of the energy of the impurity.

\begin{figure}
    \centering
    \includegraphics[width=0.49\textwidth]{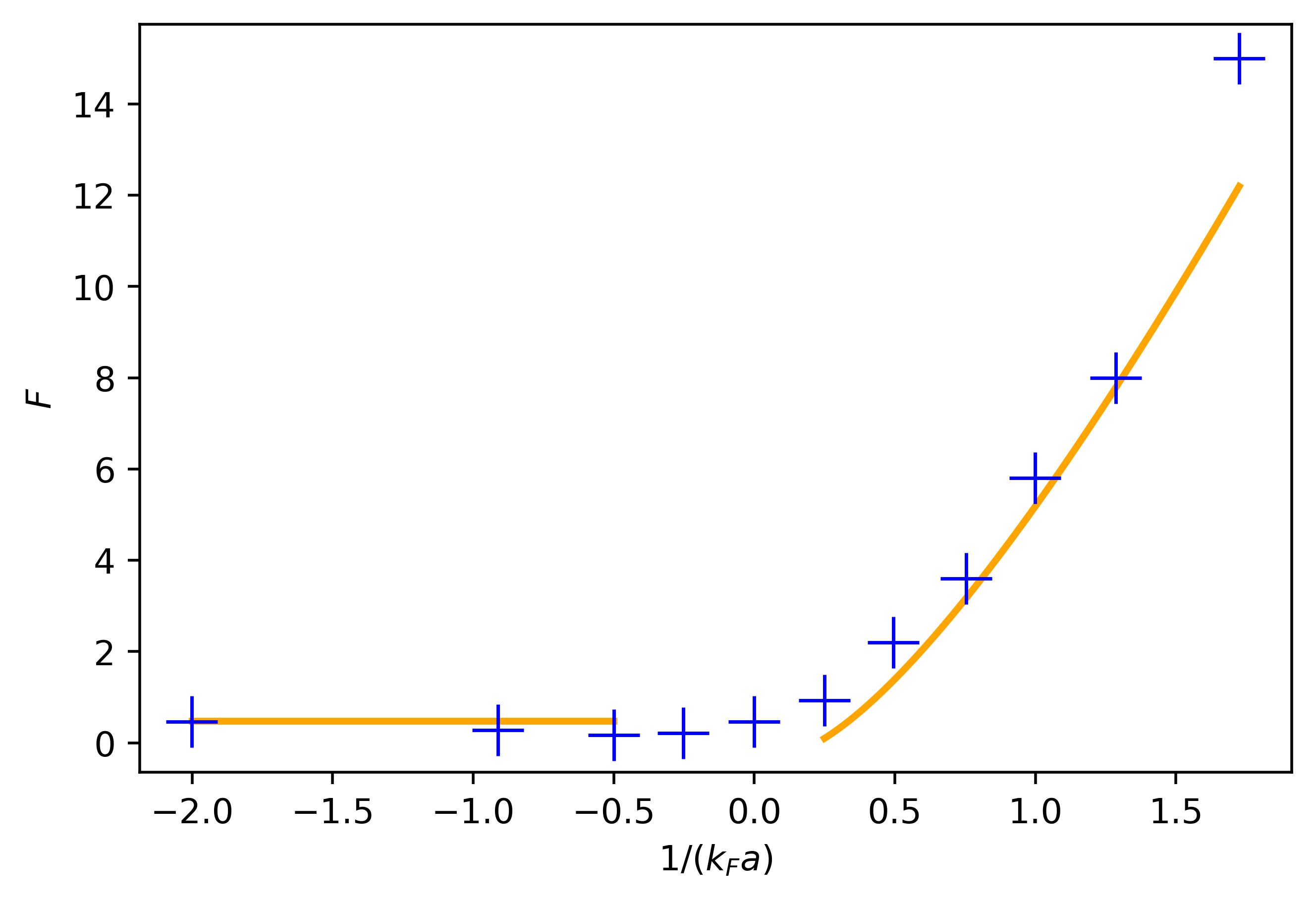}
       \includegraphics[width=0.49\textwidth]{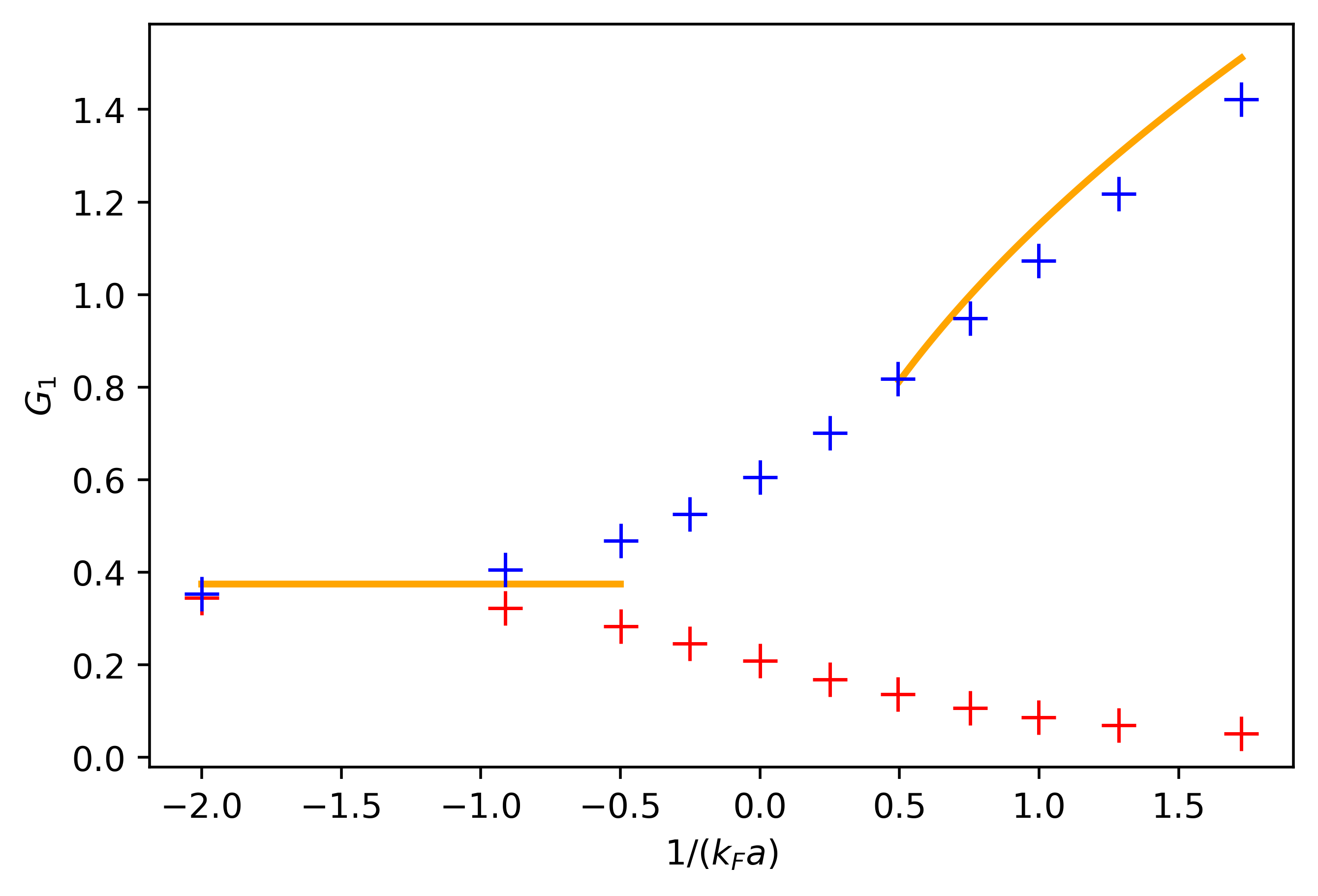}
    \caption{Blue points: Numerical values using RPA for $F(1/k_{F}a)$ (upper panel) and $G_\eta$ (lower panel) with $\eta=1$. The orange solid line correspond to the BCS and BEC asymptotic limits, and in the lower panel the red crosses correspond to the mean-field prediction accounting only for pair breaking excitations.}
\label{Fig:FRPA}
\end{figure}

Here, we solve this issue using the Random Phase Approximation (RPA) that incorporates quasi-particles interactions and recovers the collectives modes of the system (knonwn as Bogoliubov-Anderson's modes in this context) \cite{anderson1958random,minguzzi2001dynamic,Combescot2006}. In this framework, we will show that $\chi_1$ satisfies Eq. (\ref{Eq:AsymptoticChi}), with $C_2$ being given by BCS mean-field value $C_2/N=m_{\rm f}^2\Delta^2/n$ (here $\Delta$ is the superconducting gap, see Sec. \ref{sec:highq}). This allows us to regularize the polaron energy, leading to a universal function $F£$ displayed in Fig. \ref{Fig:FRPA}. The numerical values obtained using RPA coincide with the asymptotic behaviours  predicted in \cite{Pierce19few} for the BEC and BCS limits  ($1/k_F a\rightarrow \pm\infty$, orange solid lines): On the weakly attractive side, $F$ converges towards the finite limit $F(-\infty)=3/2\pi$ that can be calculated analytically for an ideal Fermi gas, while on the far BEC limit, we recover the asymptotic expansion imposed by the matching between the polaron energy with the mean-field interaction of an impurity with a Bose-Einstein condensate of deeply-bound dimers:

\be
F\left(\frac{1}{k_Fa}\right)\underset{a\rightarrow 0^+}{=}8\pi^2\kappa(\eta)\frac{m_{\rm f}}{m_{\rm r}}\frac{\ln\left(k_Fa\right)+C_{ad}}{k_Fa}+...
\ee
where $C_{ad}$ comes from the analysis of the three-body scattering problem and depends only on the mass ratio $\eta$. 
Note that the resummation scheme used in \cite{yi2015polarons} also yields the correct velue for $\kappa$ and is therefore compatible with the regularization procedure described previously. However this work incorporates quasi-particle interactions using  only a scalar response function while the RPA considers a more general $3\times 3$ response matrix that describes the coupling to both density and order parameter -- see section \ref{sec:chi} below.

Likewise, RPA can be used to calculate the second-order correction to the effective mass. Using dimensional analysis, we have

\be
\frac{\Delta m}{m_{\rm i}}=\frac{g'^2n^2}{E_F^2}G_\eta (1/k_Fa).
\label{Eq:defGeta}
\ee
The plot of $G_1$ is displayed in Fig.\ref{Fig:FRPA}, both within RPA and mean-field approximations. Both approaches coincide in the BCS limit where the excitation spectrum is dominated by pair-breaking excitations and where they asymptotically connect with the ideal gas prediction (orange line). In the BEC limit, the impurity is mostly dressed by a cloud of phonons. The BCS predictions therefore strongly underestimate the effective mass of the quasi-particle and RPA has to be used to obtain a quantitatively correct result. We see that in this regime the RPA prediction coincides with a calculation where the superfluid is described by a weakly repulsive Bose-Einstein condensate of dimers whose low-lying excitation spectrum is dominated by Bogoliubov excitations (orange line, see also Sec. \ref{sec:EffectiveMass}).

\section{Response functions in the Mean-Field and in the Random Phase Approximations}
\label{sec:chi}

We now turn to a more detailed derivation of the results sketched in the previous secction. $\chi_p$ is the central quantity giving access to both the energy and the effective mass of the polaron. To calculate it, we first note that it can be related to the density-density response of the system $\chi^R$. More precisely, let's consider the response of the superfluid to  a perturbating potential $V$ given by

$$
V=\int d^3\bm r \big(\widetilde{u}  e^{i(\bm q\cdot \bm r-\omega t)}+{\rm cc}\big)\widehat n(\bm r).
$$

In the linear regime, the density response can be written as $ \langle \hat n_q\rangle=\Omega \widetilde{u} \chi^R e^{-i\omega t}$, where, using standard perturbation theory,
\be
\begin{split}
\chi^R({\bf q},\omega)=\frac{1}{\Omega}\sum_{\alpha\neq 0}
&\Bigg[\frac{ |\langle\alpha |\hat{n}_{-{\bf q}} |0\rangle |^2}{\omega+i0^{+}-(E_{\alpha}-E_0)}\\&
-\frac{ |\langle\alpha |\hat{n}_{{\bf q}} |0\rangle |^2}{\omega+i0^{+}+E_{\alpha}-E_0}\Bigg]\label{eqdefchiRqomega},
\end{split}
\ee
and $\Omega$ is a quantization volume ensuring that $\chi^R$ is an intensive quantity.

We note that $\chi_1$ is proportional to the second term of $\chi^{R}$, and is associated with poles located on the negative part of the real axis. From this remark, we deduce that for $E>0$
\be
\chi_1({\bf q},E)=\frac{1}{n}\int_0^{+\infty}\hspace{-.3cm}d\omega' \frac{1}{E+\omega'}\bigg[
-\frac{1}{\pi}\Im(\chi^R(-{\bf q},\omega'))
\bigg]\label{eqchichiR}
\ee
The expression of $\chi_p$ for arbitrary $p$ is then obtained by diferentiation with respect to $E$ that yields
\be
\chi_p({\bf q},E)=\frac{1}{n}\int_0^{+\infty}\hspace{-.3cm}d\omega' \frac{1}{(E+\omega')^p}\bigg[
-\frac{1}{\pi}\Im(\chi^R(-{\bf q},\omega'))
\bigg]
\ee

We now turn to actual calculation of $\chi^{R}$ of an ensemble of fermions with zero range interactions characterized by a scattering length $a$. The Hamiltonian describing the system is given by

\begin{multline}
H = \int d^3\bm (\bm r)\sum_{\sigma=\updownarrow}\psi^ \dagger_\sigma(\bm r)\left(-\frac{\hbar^2}{2m_{\rm f}}\nabla^2-\mu\right) \psi^\dagger_\sigma(\bm r)\\  + g_0\int d^3\bm r \psi^\dagger_\uparrow(\bm r) \psi^\dagger_\downarrow (\bm r)\psi_\downarrow(\bm r)\psi_\uparrow(\bm r)
\end{multline}
Here $\mu$ is the chemical potential and $g_0$ is the bare coupling constant that we can relate to $g=4\pi \hbar^2 a/m_{\rm f}$ using Lippman-Schwinger's equation
\be
\frac{1}{g_0}=\frac{1}{g}-\frac{1}{\Omega}\sum_{\bm k}\frac{1}{2\varepsilon_{k}^{(\rm f)}}.
\ee

A first approach to calculate the response of the system is to consider a mean-field approximation where we replace $H$ by a quadratic Hamiltonian

\begin{multline}
H_0[U,\Delta] =\\ \int d^3\bm (\bm r)\sum_{\sigma=\updownarrow}\psi^ \dagger_\sigma(\bm r)\left(-\frac{\hbar^2}{2m_{\rm f}}\nabla^2-\mu+U(\bm r)\right) \psi^\dagger_\sigma(\bm r)\\  + \int d^3\bm r \Delta(\bm r) \psi^\dagger_\uparrow(\bm r) \psi^\dagger_\downarrow (\bm r)+\Delta^*(\bm r)\psi_\downarrow(\bm r)\psi_\uparrow(\bm r).
\end{multline}
The optimal values for the Hartree potential $U$ and the order parameter $\Delta$ are obtained variationally \cite{de2018superconductivity} and we  have  $U=g_0\langle\psi_\uparrow^\dagger\psi_\uparrow(\bm r)\rangle=g_0\langle\psi_\downarrow^\dagger\psi_\downarrow(\bm r)\rangle$ (we consider here a spin-balanced system) and $\Delta(\bm r)=g_0\langle\psi_\downarrow(\bm r)\psi_\uparrow(\bm r)\rangle$. The spectrum of the system can then be calculated using a standard self-consistent Bogoliubov transformation of the field operators and we note $\Delta_0$ and $U_0$ the ground state values of $\Delta$ and $U$.

We can now calculate $\chi^R$ by considering the response of the fermionic superfluid to the following perturbation:

\begin{multline}
H_{\rm drive} [u,\phi]= \int d^3\bm r\, u(\bm r, t)[\psi^{\dagger}_{\uparrow}(\bm r)\psi_{\uparrow}(\bm r) + \psi^{\dagger}_{\downarrow}(\bm r)\psi_{\downarrow}(\bm r)] \\ + [\phi(\bm r, t)\psi^{\dagger}_{\uparrow}(\bm r)\psi^{\dagger}_{\downarrow}(\bm r) + \text{h.c.}],
\end{multline}
where $u$ and $\phi$ are driving fields coupling respectively to the density  and the order parameter of the system.

We first consider the response  of the fermionic superfluid by describing its dynamics  using the Hamiltonian $H_0[U_0,\Delta_0]+H_{\rm drive}$, where we assume that the expressions of the Hartree potential and of the order parameter appearing in $H$ are not affected by the perturbation. As a consequence, this first approach restricts the response of the system  to the pair-breaking sector.

Following \cite{kurkjiantempereklimin2020}, we write $n=n_0+\delta n$ and $\Delta=|\Delta|e^{i\theta}\simeq \Delta_0+\delta|\Delta|+i\Delta_0 \theta$, where we chose the phase of the order parameter such that $\Delta_0$ is a real number.  Moreover, for any real physical quantity $A$ we write $A(\bm r,t)=\widetilde{A} e^{i(\bm k.\bm r-\omega t)}+\widetilde{A}^* e^{-i(\bm k.\bm r-\omega t)}$.

In the framework of linear response theory, the response of the system in terms of order parameter and density is related to the drive fields $u$,  $\phi_{+} = \Re(\phi)$ and $\phi_{-} =\Im(\phi)$ by a 3-by-3 correlation matrix $M(\bm q, \omega)$ 

\begin{equation}
\begin{pmatrix}
\Delta_0 \widetilde\theta/g_{0} \\
\delta\widetilde{|\Delta|}/g_{0} \\
\delta \widetilde n
\end{pmatrix}
=
\begin{pmatrix}
M_{11} & M_{12} & M_{13} \\
M_{21} & M_{22} & M_{23} \\
M_{31} & M_{32} & M_{33} \\
\end{pmatrix}
\begin{pmatrix}
\widetilde\phi_{-}\\
\widetilde\phi_{+} \\
2\widetilde u
\end{pmatrix}
\label{eq:def_M}
\end{equation}
where the components $M_{ij}$ can be calculated explicitly as a function of the Bogoliubov amplitudes $(u_k,v_k)$ \cite{kurkjiantempereklimin2020}\footnote{The correspondance between the matrix $M$ and the matrix $\Pi$ of \cite{kurkjiantempereklimin2020} is :
$M_{ii}=\Pi_{ii}/\Omega$ for $i=1,3$. $M_{12}=-i\,\Pi_{12}/\Omega$, $M_{21}=i\,\Pi_{12}/\Omega$, $M_{13}=(-i/2)\Pi_{13}/\Omega$, $M_{31}=(2\,i)\Pi_{13}/\Omega$, $M_{23}=(1/2)\Pi_{23}/\Omega$ and $M_{32}=2\,\Pi_{23}/\Omega$.
}. In particular,  we note that $M_{33}$ corresponds to the density-density response of the superfluid. We have indeed $  \chi_{\rm MF}^{R} = 2 M_{33}$ and applying Eq.(\ref{eqchichiR}) to $\chi_{\rm MF}^R$ yields the mean field Eq. (\ref{Eq:chipMF}) for $\chi_p$.

The perturbation of the order parameter and of the Hartree potential contradicts the assumption that $U$ and $\Delta$ are not modified in the expression of $H_0$. In the Random Phase Approximation, we solve this contradiction by considering a self-consistent response of the system to $H_{\rm drive}$. In other words, we now describes the system by $H_{\rm RPA}=H_0[U+\delta U,\Delta+\delta\Delta]+H_{\rm drive}[u,\phi]$.

We note that, by construction, $H_0[U_0+\delta U,\Delta_0+\delta\Delta]+H_{\rm drive}[u,\phi]=H_0[U_0,\Delta_0]+H_{\rm drive}[u+\delta U,\phi+\delta\Delta]$, from which we conclude that the response of the system can be described using the previous calculation, but considering now an effective self-consistent drive defined by

\be
\begin{pmatrix}
\widetilde \phi_{-}\\
\widetilde\phi_{+} \\
2\widetilde u
\end{pmatrix}_{\rm eff}
=
\begin{pmatrix}
\widetilde\phi_{-}\\
\widetilde\phi_{+} \\
2\widetilde u
\end{pmatrix}
+
\begin{pmatrix}
\Delta_0\widetilde\theta\\
\delta|\widetilde\Delta|\\
g_0\delta\widetilde n
\end{pmatrix}
\ee

Using the non self-consistent  approach, we have

\be
\begin{pmatrix}
\Delta_0 \widetilde\theta \\
\delta |\widetilde\Delta| \\
g_{0}\delta\widetilde n
\end{pmatrix}
=
g_0M
\begin{pmatrix}
\widetilde\phi_{-}\\
\widetilde\phi_{+} \\
2\widetilde u
\end{pmatrix}_{\rm eff}
\ee
hence, using the definition of the effective driving fields,
\be
\begin{pmatrix}
\Delta_0 \widetilde\theta \\
\delta |\widetilde\Delta| \\
g_{0}\delta\widetilde n
\end{pmatrix}
=
g_0M_{\rm RPA}
\begin{pmatrix}
\widetilde\phi_{-}\\
\widetilde\phi_{+} \\
2\widetilde u
\end{pmatrix}
\ee
with

\begin{equation}
M_{\rm RPA} = \frac{M}{\mathbb{I}-g_{0} M}
\end{equation}

The compressibility of the system corresponds to $\chi_{\rm RPA}^R=2\left(M_{\rm RPA}\right)_{33}$. In the zero-range limit where $g_{0} \rightarrow 0$, the diagonal terms $M_{ii}$ with $i=1,2$ need to be renormalized to obtain finite values. Introducing  $\widetilde{M}_{ii} = M_{ii} - 1/g_{0} $, the compressibility can be written as a sum of a mean-field term and an interaction-induced term proportional to $\Delta^{2}$  \cite{Combescot2006}:

\begin{equation}
\chi_{\rm RPA}^{R} = 2\frac{\begin{vmatrix}
\widetilde{M}_{11} & M_{12} & M_{13} \\
M_{21} & \widetilde{M}_{22} & M_{23} \\
M_{31} & M_{32} & M_{33} \\
\end{vmatrix}}{\begin{vmatrix}
\widetilde{M}_{11} & M_{12} \\
M_{21} & \widetilde{M}_{22} \\
\end{vmatrix}}
= \chi_{\rm MF}^{R} + \chi_{\rm int}^{R}
\end{equation}

with

\begin{equation}
\chi_{\rm int}^{R} = 2\frac{-M_{23}M_{32}\widetilde{M}_{11} - M_{13}M_{31}\widetilde{M}_{22} + 2M_{12}M_{23}M_{31}}{\widetilde{M}_{11}\widetilde{M}_{22}-M_{12}M_{21}}\label{Eq:chiint}
\end{equation}

We note that, all matrix elements being calculated in the BCS superfluid, their imaginary part is non-zero only if the frequency $\omega$ is greater than 2$\Delta$ ; however, the zeros of the denominator of the beyond-mean-field term yield a non-zero imaginary part for frequencies below 2$\Delta$. In other terms, in addition to the contribution of the pair-breaking continuum, the RPA gives us both the dispersion relation of the collective modes and their contribution to the density-density response function. These collective modes were shown to turn into the usual Bogoliubov condensate modes in the BEC limit \cite{Combescot2006}, and they become negligible in the BCS limit due to the exponential decrease of the gap.

\begin{figure}
    \centering
    \includegraphics[width=0.49\textwidth]{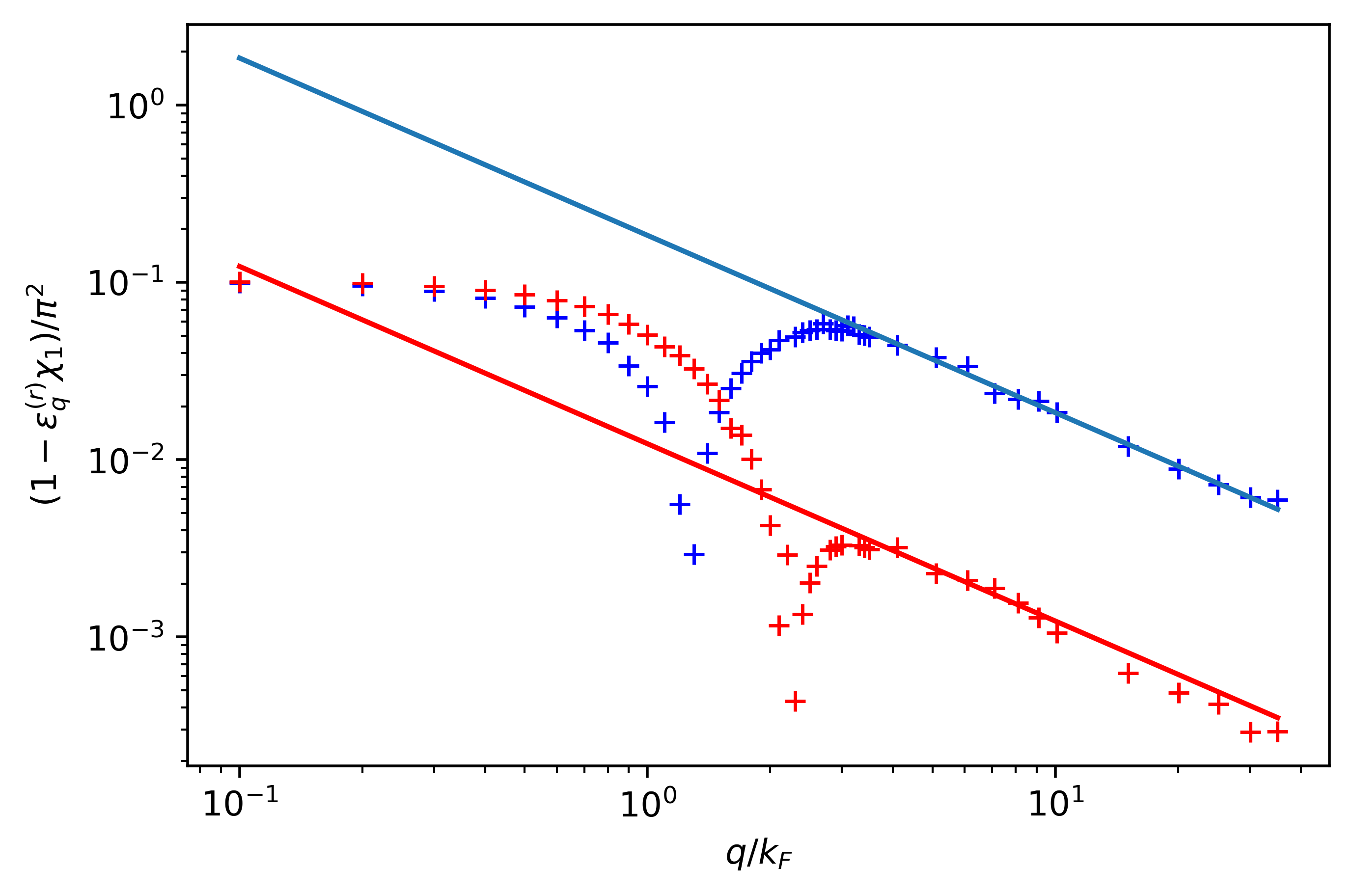}
    \caption{Log-log representation of the mean-field (red) and the RPA (blue) numerical absolute values of the subdominant terms in $\chi_{1}(\bm{q}, \varepsilon_{q}^{{\rm i}})$ rescaled by the dominant $O(q^{-2})$ term for a mass ration $\eta=1$. The solid lines represent the $1/q^3$ asymptotic behaviors for mean-field (red) and RPA (blue) associated with their respective values of $\kappa$ (Eqs. (\ref{Eq:kappaMF})  and (\ref{eq:Kappa}) ). We see that the numerics agree with the predicted asymptotic behaviours.  We see that, contrary to the MF result, the RPA result satisfies the asymptotic behaviour required by renormalization of the polaron energy.}
    \label{fig:chi}
\end{figure}

We calculate numerically $\chi^R$ and $\chi_p$ using the approach laid out in \cite{kurkjiantempereklimin2020}, At large $q$, both the mean-field and RPA results behave as $1/q^2$ and we display in Fig. \ref{fig:chi} the sub-leading contribution for a unitary Fermi gas. As discussed in the introduction, we observe that the mean-field prediction underestimates the value of $\chi_1$ while the RPA approach yields the asymptotic behavior given in Eq. (\ref{Eq:AsymptoticChi}) and is therefore compatible with a renormalization  using three-body collisions.

\section{High-$q$ behavior of $\chi_1(q,q^2/(2m_{\rm i}))$}
\label{sec:highq}
Here, we  prove analytically that, as revealed numerically in previous section,  $\chi_{1}$ follows  the asymptotic behavior (\ref{Eq:AsymptoticChi}). In section \ref{sec:highqMFq3}, we first prove that the dominant $O(q^{-2})$ term comes from $\chi^R_{\rm MF}$, the mean-field contribution to the density-density response function
and that the mean-field contribution gives a $O(q^{-3})$ term in (\ref{Eq:AsymptoticChi}), with a constant $\kappa_{\rm MF}$ given by the first two terms in (\ref{eq:Kappa}). Finally, in section \ref{sec:highqnonMFq3}, we prove that $\chi^R_{\rm int}$, the non mean-field contribution to the density-density response function gives  a $O(q^{-3})$ contribution to $\chi_1(q,q^2/(2m_{\rm i}))$ and the third term in equation (\ref{eq:Kappa}) for $\kappa$.

\subsection{Subdominant $O(q^{-3})$ contribution: MF term}
\label{sec:highqMFq3}
As mentionned in the first section, the $1/q^2$ behaviour of $\chi_1$
 originates from the high energy response of the system that corresponds to free particle excitations. As a consequence, the leading UV behaviour of $\chi$ is fully contained in the BCS mean-field term $\chi_{\rm MF}$ (see Eq. (\ref{Eq:HighQExpansion}-\ref{Eq:AsymptoticChi}).
In order to reveal the sub-leading  $q^{-3}$ dependence,  we subtract the term of order $q^{-2}$ to $\chi_{\rm MF,\,1}$ and we have
\begin{equation}
\begin{split}
 \chi_{\rm MF,\,1}(q,\frac{q^2}{2 m_{\rm i}})&-\frac{2\,m_{\rm r}}{q^2}\\&=
 \int\hspace{-.2cm}\frac{d^3k}{(2\pi)^3}\bigg[
\frac{(E_+E_--\xi_+\xi_-+\Delta^2)}{2E_+E_-\big(\frac{q^2}{2 m_{\rm i}}+E_++E_-\big)
 }\\
& -\frac{2 m_{\rm r}}{q^2}\bigg(\frac{E_+-\xi_+}{2 E_+}+\frac{E_--\xi_-}{2 E_-}\bigg)
\bigg]\frac{1}{n}
\label{eqchitildeMFsubstr}
 \end{split}
\end{equation}
where we have used the notations $E_{\pm}=E_{k_{\pm}}$ with ${\bf k}_{\pm}={\bf k}\pm {\bf q}/2$ and $\xi_{\pm}=\xi_{k_{\pm}}$ and $E_k=\sqrt{\xi_k^2+\Delta^2}$, $\xi_k=k^2/(2m_{\rm f})-\mu$. To obtain Eq.(\ref{eqchitildeMFsubstr}), we have furthermore used that $n/2=\int\hspace{-.2cm}\frac{d^3k}{(2\pi)^3}\big(\frac{E_+-\xi_+}{2 E_+}\big)=\int\hspace{-.2cm}\frac{d^3k}{(2\pi)^3}\big(\frac{E_--\xi_-}{2 E_-}\big)$.
In the $q\to\infty$ limit, we define a cutoff $\Lambda$ such that $$q\gg\Lambda\gg k_{low},$$ where
$k_{low}$ is a low energy wavevector scale defined by $k_{low}=\max\big(\sqrt{2m_{\rm f}\Delta},\sqrt{2m_{\rm f}|\mu|}\big)$.
In Appendix \ref{appendixrestqMFq3}, we show that that the contribution from the  domains $\{k_{\pm }<\Lambda\}$ (\ref{eqchitildeMFsubstr}) are negligible at leading order.
 In the domain $\{k_{\pm}>\Lambda\}$, we can replace at lowest order
$(E_{\pm}-\xi_{\pm})/E_{\pm}$ by $2 m_{\rm f}^2\Delta^2/k_{\pm}^4$,
$E_+E_--\xi_+\xi_-+\Delta^2$ by $\frac{1}{2}\Delta^2\big(\frac{k_+^2}{k_-^2}+\frac{k_-^2}{k_+^2}+2\big)$ and $E_{\pm}$ by $k_{\pm}^2/(2m_{\rm f})$ in the denominators. The only remaining scale is $q$ and we rescale ${\bf k}$ by $q$.
It is easily seen that the integral converges in $+\infty$ and near $\widetilde{k}_{\pm}=0$,  due to the first term in the integrand of (\ref{eqchitildeMFsubstr})\footnote{
Indeed,  close to $\widetilde{k}_-=0$ for instance,  the $k^{-4}$ contributions of the two terms in (\ref{eqchitildeMFsubstr}) cancel.
The next order term is proportionnal to ${\bf\widetilde{ k}}_-\cdot\hat{q}/k_-^4$ and vanishes also after angular integration.
The lowest order term is then of order $k_-^{-2}$.  After multiplication by the $k_-^{2}$ of the jacobian of spherical coordinates,  it tends to a constant.
This shows that the integral \ref{eqJMF} is convergent.}.
The contribution of order $q^{-3}$ is
\be
\frac{1}{n}\frac{m_{\rm r}\,m_{\rm f}^2\,\Delta^2}{q^3} J_{\rm MF}(\eta)
\ee
where
\be
J_{\rm MF}(\eta)= \idk\bigg[
\frac{
\big(\frac{1}{k_-^2}+\frac{1}{k_+^2}\big)^2
}{
\frac{1}{1+\eta}+\frac{\eta}{1+\eta}(2 k^2+\frac{1}{2})
}
-\frac{1}{k_-^4}-\frac{1}{k_+^4}
\bigg]\label{eqJMF}
\ee

$J_{\rm MF}$ can be evaluated analytically and we find
$$J_{\rm MF}(\eta)=\pi^2\frac{\eta+1}{\eta}\big(
\eta\arctan{\frac{1}{\sqrt{\eta(\eta+2)}}}
-\frac{\sqrt{\eta^3(\eta+2)}}{(\eta+1)^2}\big).$$

If we add the contributions of order $q^{-2}$ and $q^{-3}$, we find the following high-$q$ expansion for $\chi_{\rm MF,\,1}(q,\frac{q^2}{2 m_{\rm i}})$
\be
\begin{split}
\chi_{\rm MF,\,1}(q,\frac{q^2}{2 m_{\rm i}})=\frac{2\,m_{\rm r}}{q^2}
 \bigg(1-\pi^2\,\kappa_{\rm MF}(\eta)\frac{m_{\rm f}}{m_{\rm r}}
 \left(
 \frac{m_{\rm f}^2\,\Delta^2}{nq}\right)
 \bigg)\\+o(q^{-3})
 \end{split}
\ee
where
\be
\kappa_{\rm MF}(\eta)=\frac{\sqrt{\eta^3(\eta+2)}}{2\pi^3(\eta+1)^2}-\frac{\eta}{2\pi^3}\arctan{\frac{1}{\sqrt{\eta(\eta+2)}}}
\label{Eq:kappaMF}
\ee
In this expression,  $m_{\rm f}^2\Delta^2$ is  Tan's contact per unit volume
$C_2/\Omega$
in a mean-field BCS theory.  The ratio $\frac{m_{\rm f}^2\,\Delta^2}{n}$ is therefore equal to $C_2/N$,  where $N$ is the total (summed on spins) number of particles.  We recover a result similar to the Eq.(\ref{Eq:AsymptoticChi}),  but with $\kappa_{\rm MF}$ instead of the full $\kappa$.  The non-MF part will complete the value of $\kappa$,  as we will see in the next section.

\subsection{Subdominant $O(q^{-3})$ contribution: non MF term}
\label{sec:highqnonMFq3}
We find the high-$q$ behaviour of the non-mean field contribution of the response function $\chi^R_{\rm int}(q,\omega)$ in (\ref{Eq:chiint}) by taking the ratio $\widetilde{\omega}=\omega/(q^2/(2m_{\rm f}))$ fixed. This means that in the $q\to \infty$ limit, the frequency $\omega$ also tends to infinity. We find that in this regime, $\widetilde{M}_{11}(q,\omega)\sim m_{\rm f}\,q\,f_{11}(\widetilde{\omega}+i\,0^+)$, $\widetilde{M}_{22}(q,\omega)\sim m_{\rm f}\,q\,f_{22}(\widetilde{\omega}+i\,0^+)$,
$M_{12}(q,\omega)\sim (-i)m_{\rm f}\,q\,f_{12}(\widetilde{\omega}+i\,0^+)$,
$M_{13}(q,\omega)\sim (-i/2)(\Delta\,m_{\rm f}^2\,/q)\,f_{13}(\widetilde{\omega}+i\,0^+)$ and
$M_{23}(q,\omega)\sim (1/2)(\Delta\,m_{\rm f}^2\,/q)\,f_{23}(\widetilde{\omega}+i\,0^+)$.
These scaling behaviors are easily found using the expressions of the $\Pi_{ij}$'s of \cite{kurkjiantempereklimin2020} and the correspondance between $\Pi_{ij}$ and $M_{ij}$.
Let us consider for instance $M_{23}$. From equations (36) and (37)  of \cite{kurkjiantempereklimin2020}, we have
$M_{23}(q,\omega)=-\frac{1}{2}\idk \frac{\epsilon^+_{{\bf k}\,{\bf q}}w^+_{{\bf k}\,{\bf q}}W^-_{{\bf k}\,{\bf q}}}{\omega^2-(\epsilon^+_{{\bf k}\,{\bf q}})^2}$, with $\epsilon^+_{{\bf k}\,{\bf q}}=E_++E_-$ and $w^+_{{\bf k}\,{\bf q}}W^-_{{\bf k}\,{\bf q}}=\frac{\Delta}{2\,E_+\,E_-}(\xi_++\xi_-)$.
We separate the ${\bf k}$ space into two parts: $k<\Lambda$ and $k>\Lambda$, where the cutoff $\Lambda$ fulfills $k_{\rm low}\ll \Lambda\ll q$.
For $k<\Lambda$, at lowest order, the integrand is of order $\Delta/(q^2/m_{\rm f})^2$. Indeed, we can neglect $k$ compared to $q$ and $\Delta$ and $\mu$ compared to $q^2/m_{\rm f}$. The integration gives a volume factor of the order $\Lambda^3$ and the contribution to $M_{23}$ is of order $\Delta\Lambda^3/(q^2/m_{\rm f})^2$.
Compared to the $q^{-1}$ contribution, of the $k>\Lambda$ domain this is negligible. Indeed, we find a ratio of order $\Lambda^3/q^3\ll 1$. In the $k>\Lambda$ domain, we rescale $k$ by $q$ and we can neglect $\Delta$ and $\mu$ in $\xi_{\pm}$ and $E_{\pm}$.
The integration volume gives a factor $q^3$ and the integrand a factor $\Delta /(q^2/(2m_{\rm f}))^2$. This gives a $q^{-3}$ dependance.  We find
$f_{23}=-2\idk\frac{((k_+)^{-2}+(k_-)^{-2})(2k^2+\frac{1}{2})}{(\widetilde{\omega}+i0^+)^2-(2k^2+\frac{1}{2})^2}$. In the same manner, we find
$f_{13}=-2\idk\frac{((k_+)^{-2}+(k_-)^{-2})(\widetilde{\omega})}{(\widetilde{\omega}+i0^+)^2-(2k^2+\frac{1}{2})^2}$,
$f_{12}=2\idk\frac{\widetilde{\omega}}{(\widetilde{\omega}+i0^+)^2-(2k^2+\frac{1}{2})^2}$ and
$f_{11}=f_{22}=2\idk\big(\frac{2k^2+\frac {1}{2}}{(\widetilde{\omega}+i0^+)^2-(2k^2+\frac{1}{2})^2}+\frac{1}{2k^2}\big)$.
If we use the asymptotic expressions of the $M_{ij}$'s in (\ref{Eq:chiint}), we find $\chi^R_{\rm int}\sim \frac{\Delta^2\,m_{\rm f}^3}{q^3}\frac{2\,f_{12}\,f_{23}\,f_{13}-f_{11}(f_{13}^2+f_{23}^2)}{(f_{11}+f_{12})(f_{11}-f_{12})}$.

 Finally, we use (\ref{eqchichiR}) to determine the large $q$ behavior of $\chi_1(q,q^2/(2m_{\rm i}))$.

 $$
 \chi_1(q,\frac{q^2}{2\,m_{\rm i}})\sim\frac{\Delta^2 m^3}{n\,q^3}4\sqrt{2}
 \Re\big(
 \int_{\frac{1}{2}}^{+\infty}\frac{X_+^2}{(\widetilde{\omega}+\frac{1}{\eta})\sqrt{\widetilde{\omega}-\frac{1}{2}}}d\widetilde{\omega}
 \big)
 ,$$
 where $X_{+}=f_{23}+ f_{13}=-2\idk\frac{((k_+)^{-2}+(k_-)^{-2})}{(\widetilde{\omega}+i0^+)-(2k^2+\frac{1}{2})}$.
 In the last step, we use the integral expression for $X_+$ and write $X_+^2$ as a double integral on wave vectors and exchange the order of integrations (we integrate on $\widetilde{\omega}$ first). After integration on the frequency $\widetilde{\omega}$, the integrals on the wave vectors factorize. The integral we need to calculate is  $\idk\frac{((k_+)^{-2}+(k_-)^{-2})}{2\,k^2+\frac{1}{2}+\frac{1}{\eta}}$. It is found to be equal to $\frac{1}{2\pi}\arctan\sqrt{\frac{\eta}{\eta+2}}$. As a result we find for the high $q$ behavior of the non mean field part of $\chi_1$
 \begin{equation}
 \begin{split}
 \chi_{1,non-MF}(q,\frac{q^2}{2m_{\rm i}})\sim
 \frac{(m_{\rm f}\Delta)^2\,m_{\rm f}}{n}\frac{8}{\pi}\sqrt{\frac{\eta}{\eta+2}}\\
 \times\bigg(\arctan\sqrt{\frac{\eta}{\eta+2}}\bigg)^2\frac{1}{q^3}
 \end{split}
 \end{equation}
As mentionned before,  $(m_{\rm f}\Delta)^2$ is Tan's contact per unit volume, and we see from (\ref{Eq:AsymptoticChi}) that the non mean-field contribution gives the third contribution to $\kappa$ in (\ref{eq:Kappa}).


\section{BCS and BEC limits for the effective mass}
\label{sec:EffectiveMass}

The expressions of $\chi^{R}$ found using RPA and the mean-field approximations can also be used to obtain the effective mass of the polaron. As explained in the end of Sec.I, we can write the second-order perturbation to the effective mass $m^{*}$ in units of the impurity mass $m_{\rm i}$: it takes the form of a mean-field energy in units of the Fermi energy $g'n/E_{F}$ squared times a dimensionless function $G_{\eta}(1/k_{F}a)$ (Eq.(\ref{Eq:defGeta})).
Along the crossover, Eq.(\ref{Eq:EffectiveMass}) gives us a relation between the response function $\chi_3(\bm q,\varepsilon_q^{(\rm i)})$ and the correction to the effective mass $G_{\eta}$:

\begin{eqnarray}
G_{\eta} = \frac{4}{3}\frac{E_{F}^{2}}{n}\int\frac{d^3\bm q}{(2\pi)^3}\varepsilon_q^{(\rm i)}\chi_3(\bm q,\varepsilon_q^{(\rm i)})
\end{eqnarray}

As mentionned in the introductory part, we see in Fig. \ref{Fig:FRPA} that the mean-field and RPA results are consistent in the BCS limit (that is, when $1/k_{F}a$ tends to $-\infty$).  In this regime, the bath is indeed made of two non-interacting Fermi seas of opposite spins and its elementary excitations are particle-hole pairs which are correctly captured within the BCS approximation, as well as its RPA extension. $\chi_{3}$ is then given by

\be
\chi_3(q,E)=\frac{2}{n}\int_{\substack{{k<k_F}\\{|\bm k+\bm q|>k_F}}}\hspace{-.2cm}\frac{d^3k}{(2\pi)^3}\frac{1}{\left(E+q^2/2m_{\rm f}+\bm k\cdot\bm q/m_{\rm f} \right)^3},
\label{Eq:chi3MF}
\ee
In particular, in the equal-mass case, an analytical calculation yields $G_1(-\infty)=3/8$ that coincides with the numerical results displayed in Fig. (\ref{Fig:FRPA}).

In the opposite limit, $1/(k_{F}a)\rightarrow +\infty$, the mean-field approximation strongly underestimates the correction to the effective mass. Indeed, in this regime  the fermions become so tightly bound that Cooper pairs lose their internal degrees of freedom, and we can treat the system as a condensate of composite bosons \cite{nozieres1985bose, Leggett81}. Pair breaking excitations are therefore suppressd and We can apply Eq.(2) to an ensemble of bosons, provided we replace $\chi_{3}$ by the response function  $\chi_{3}^{b}$ of the condensate, calculated within the mean-field Bogoliubov approximation. The fermionic physical parameters $n$, $m_{{\rm f}}$, $a$ and $\mu$ are replaced by their bosonic equivalents: the boson density $n_{{\rm b}}=n/2$, the boson mass $m_{{\rm b}}=2m_{{\rm f}}$ and the boson-boson scattering length (in the framework of BCS theory) $a_{{\rm bb}} = 2 a$ which yields us the bosonic chemical potential $\mu_{{\rm b}} = \frac{4\pi a_{{\rm bb}}}{m_{{\rm b}}}n_{{\rm b}} = gn/2$. We also need to replace the impurity-fermion coupling constant $g'$ by its impurity-boson counterpart given by Born's approximation $g'_{{\rm b}} = 2g'$.
Eq.(2) transforms into

\begin{eqnarray}
 \frac{1}{m^*}&=&\frac{1}{m_{\rm i}}\left[1-\frac{8 g'^2n}{3}\int\frac{d^3\bm q}{(2\pi)^3}\varepsilon_q^{(\rm i)}\chi_3^{b}(\bm q,\varepsilon_q^{(\rm i)})\right]
\end{eqnarray}
with

\be
\chi_{3}^{{\rm b}}(q,E)=\frac{\left |u^{{\rm b}}_{\bm q} + v^{{\rm b}}_{\bm q}\right |^2}{\left(E+ E^{{\rm b}}_{\bm q}\right)^3}
\label{Eq:chi3b}
\ee
with $u^{{\rm b}}_{\bm q}$ and $v^{{\rm b}}_{\bm q}$ being the amplitudes of the Bogoliubov modes of the condensate and $E^{{\rm b}}_{\bm q}$ their energy.

In this integral, energies and wave-vectors can be rescaled by the  chemical potential and the healing length of the bosonic dimers. We then readily find that $G_{1}(1/k_{F}a)$ scales like the square root of $1/k_{F}a$ with

\begin{eqnarray}
G_{\eta}\left(\frac{1}{k_{F}a}\right) \underset{a\rightarrow0^{+}}{=}\sqrt{6\pi}I(\eta)\sqrt{\frac{1}{k_{F}a}}
\end{eqnarray}
where

\begin{eqnarray}
I(\eta) &=& \int\limits_{0}^{+\infty}\frac{q^{4}dq}{\eta}\sqrt{\frac{\frac{q^{2}}{2}}{\frac{q^{2}}{2}+1}}\left(\frac{q^{2}}{\eta} +  \sqrt{\frac{q^{2}}{2}\left(\frac{q^{2}}{2} + 1\right)}\right)^{-3},
\end{eqnarray}
with $I(1)\simeq 0.265$. As shown in Fig. (\ref{Fig:FRPA}), this asymptotic behaviour coincides with the numerical results based on the RPA approximation.

\section{Conclusion and outlook}

A possible way to test this prediction for the energy shift $\Delta E$ is to use radio-frequency spectroscopy which provided a very sensitive probe for the study of Fermi \cite{schirotzek2009ofp,kohstall2011metastability,amico2018time,yan2019boiling} and Bose \cite{jorgensen2016observation,hu2016bose,yan2020bose} polarons, However, as described below, signatures of beyond mean-field effects might be already observable in the measurement  of the oscillation frequency of an impurity of $^7$Li inside a superfluid of $^6$Li reported in  \cite{Ferrier2014Mixture}.

To model this experiment,  we  assume that  the impurity and the background superfuid are trapped by the same potential $V$. We can then describe the semi-classical dynamics of the polaron using the Hamiltonian

\be
h(\bm r,\bm p)=\frac{p^2}{2m^*}+V(\bm r)+\Delta E(n(\bm r)).
\ee

Within LDA, the density profile of the fermionic background can be obtained from its equation of state $n(\mu)$ using the prescription $\mu(\bm r)=\mu_0-V(\bm r)$, where $\mu_0$ is the global chemical potential of the fermions.

For vanishingly small amplitude oscillations of the impurity, we have $V(\bm r)\ll \mu_0$ and the energy of the quasi-particle can be approximated by

\be
h(\bm r,\bm p)=\Delta E(n_0)+\frac{p^2}{2m^*}+\left(1+\left.\frac{d \Delta E}{dn}\frac{dn}{d\mu}\right|_0\right)V(\bm r),
\ee
where the subscript 0 means that the quantity is evaluated at the center of the fermionic cloud.

If $V$ is a harmonic potential characterized by a frequency $\omega$, the effective oscillation frequency of the impurity is now
\be
\omega^*=\omega\sqrt{\frac{m_{\rm i}}{m^*}\left(1+\left.\frac{d \Delta E}{dn}\frac{dn}{d\mu}\right|_0\right)}.
\ee

If we treat the impurity-bath interaction in a mean-field approximation (in other terms, if we restrain ourselves to the first-order term in the perturbative expansions in $g'$), the relative frequency correction $(\omega^{*(1)}-\omega)/\omega$ is given by $\frac{g'}{2}\left.\frac{dn}{d\mu}\right|_0$.

Consequently, if the beyond-mean-field relative frequency correction $\beta$ is small, we can write it as the sum of the contributions of the second-order term in $g'$ in the ground-state energy and in the effective mass respectively:

\be
\frac{\delta\omega/\omega - \delta\omega^{(1)}/\omega}{\delta\omega^{(1)}/\omega} \approx \beta^{(2)}_{GS} + \beta^{(2)}_{m^{*}}
\ee

At unitarity, we get

\be
\frac{\delta\omega^{(1)}}{\omega} = \frac{g' m_{\rm f}k_{F}}{2\hbar^{2}\pi^{2}\xi}
\ee

 and

\be
\beta^{(2)}_{GS} = \frac{4}{3}k_{F}a'\left[F(0) - 2\pi\kappa(\eta)\frac{m_{\rm f}}{m_{\rm r}}\frac{C_{2}}{Nk_{F}}\left(\ln(k_{F}R_{3}) + \frac{1}{4}\right)\right]
\ee

We will here compute the value of these second-order frequency correction in the conditions reported in \cite{Ferrier2014Mixture}, namely $k_{F}a' = 10^{-2}$, $R_{3}=1.50 a'$ and $m_{\rm i} = \frac{7}{6}m_{\rm f}$. Knowing that at unitarity the equation of state of the superfluid is the same as the one of an ideal Fermi gas up to a multiplying constant $\xi=0.376$ \cite{zwerger2012BCSBEC,forbes2011resonantly,carlson2011auxiliary,endres2013lattice,luo2009thermodynamic,navon2010equation,ku2012revealing}, we obtain that the second term in the above expression (which we may call the three-body term) gives a relative correction about $-4.7\%$, whereas the two-body term (the one involving $F$) can be estimated to be around $1.1\%$.

Along the crossover, the correction due to the effective mass $\beta^{(2)}_{m^{*}}$ reads

\be
\beta^{(2)}_{m^{*}} = -\xi G_{\eta}(0)\frac{m_{\rm f}}{m_{\rm r}}  \frac{k_{F}a'}{9\pi}
\ee
\\
which would give us a value of about $-0.02\%$.

All these terms being second-order in $g'$, their relative values at unitarity are fixed: the effective mass will always (in the RPA and as long as $g'$ is small) have an effect on the frequency correction that is two orders of magnitude smaller than the two-body term of the ground-state energy and about 250 times smaller than the three-body term.

Interestingly, \cite{Ferrier2014Mixture} reported a $\sim 14$~\% upshift of the mean-field prediction with respect to the  measured oscillation frequency at unitarity. This discrepancy is slightly beyond the 10\% experimental error bar and the beyond-mean-field correction calculated in the present paper lowers this difference to only $\sim 10$~\% that is now within the experimental uncertainty.

In this estimate of the beyond-meand field contribution to the frequency shift, we have neglected the drag between the two isotopes due to  Andreev-Bashkin's effect \cite{andreev1976three} that was predicted first in nuclear physics and has recently been the focus of several theoterical studies in the context of ultracold gases \cite{nespolo2017andreev,parisi2018spin,hossain2021detecting}. The study of the interplay between these two effects is beyond the scope of this article and will be addressed in future work.


\begin{acknowledgments}
We thank Matthieu Pierce for early contributions to this work as well as Félix Werner, Hadrien Kurkjian, Christophe Salomon, Georg Bruun, and Ragheed Alhyder for stimulating discussions. This work was supported by European Union (ERC grant CritiSup2), CNRS (80Prime project TraDisQ1D) and DIM SIRTEQ (1DFG).
\end{acknowledgments}

\appendix
\section{Justification of neglected terms in the high-$q$ limit of $\chi_1(q,q^2/(2\,m_i))$}
\label{appendixrestqMFq3}
We now briefly justify why the integrations in domains $\{k_{-}<\Lambda\}$ and $\{k_{+}<\Lambda\}$ in (\ref{eqchitildeMFsubstr}) are negligible at this order.
In the domain $\{k_{-}<\Lambda\}$,  we take the limit $q\to+\infty$ and keep ${\bf k}_-$ finite.
We easily find
$\frac{E_+-\xi_+}{4 E_+}=\frac{m_{\rm f}^2\Delta^2}{2\,q^4}+O(q^{-5})$,
$\frac{E_+E_--\xi_+\xi_-+\Delta^2}{2E_+E_-}=
\frac{E_--\xi_-}{2 E_-}+\frac{\Delta^2}{E_-\xi_+}+O(\xi_+^{-3})$,
$\frac{1}{\xi_+}=\frac{2m_{\rm f}}{q^2}(1-\frac{2\hat{q}\cdot{\bf k}_-}{q}+\frac{4(\hat{q}\cdot{\bf k}_-)^2-2m_{\rm f}\xi_-}{q^2}+O(q^{-3}))$,
$\frac{1}{\frac{q^2}{2 m_{\rm i}}+E_-+E_+}=\frac{2m_r}{q^2}(1-\frac{2m_r}{m_{\rm f}}\frac{\hat{q}\cdot{\bf k}_-}{q}+\frac{4(\frac{m_r}{m_{\rm f}})^2(\hat{q}\cdot{\bf k}_-)^2-2m_r(\xi_-+E_-)}{q^2}+O(q^{-3}))$.
  Using these results,  we find for the expansion of the integrand of (\ref{eqchitildeMFsubstr}) in the domain $1$ (the terms of order $q^{-2}$ compensate)
  $$
  \frac{A}{q^3}+\frac{B}{q^4}+o(q^{-5})
  $$
  where $A=-\frac {1}{n}\bigg( \frac{2m_{\rm r}^2}{m_{\rm f}}\frac{(E_--\xi_-)}{E_-}\hat{q}\cdot{\bf k}_-\bigg)$ and
 $
 B=\frac{2m_{\rm r}}{n}\bigg(
  \frac{(E_--\xi_-)}{2 E_-}\big(4(\frac{m_{\rm r}}{m_{\rm f}})^2(\hat{q}\cdot{\bf k}_-)^2-2 m_r(\xi_-+E_-)\big)+\frac{m_{\rm f}\Delta^2}{E_-}
  \bigg)
 $

After angular integration, the first term of order $q^{-3}$ vanishes.  The term of order $q^{-4}$ behaves as $k_-^{-2}$ if $k_-\gg k_{low}$.  After integration on ${\bf k}_-$ ($k_-<\Lambda$),  it gives a contribution
of order $\Lambda/q^4$.  Compared to the term of order $q^{-3}$, we find a ratio $\Lambda/q$. Therefore it is negligible.  In the same way,  we find the integration in domain $2$ is negligible.

\bibliography{bibliographie}
\end{document}